\documentstyle[epsfig]{mn}{}
\begin{document}

\def\mpc{h^{-1} {\rm{Mpc}}}
\def\up{h^{-3} {\rm{Mpc^3}}}
\def\uk{h \ {\rm{Mpc^{-1}}}}
\def\lsim{\mathrel{\hbox{\rlap{\hbox{\lower4pt\hbox{$\sim$}}}\hbox{$<$}}}}
\def\gsim{\mathrel{\hbox{\rlap{\hbox{\lower4pt\hbox{$\sim$}}}\hbox{$>$}}}}
\def\kms {\rm{km~s^{-1}}}
\def\apj {ApJ}
\def\aj {AJ}
\def\aa {A\&A}
\def\mnras {MNRAS}

\title{Galaxy groups in the 2dF galaxy redshift survey: \\
Large Scale Structure with Groups}

\author[A. Zandivarez, M. Merch\'an \& N. Padilla]
{Ariel Zandivarez$^{1,2}$, Manuel E. Merch\'an $^{1,2}$ \& Nelson D. Padilla$^{3}$\\
$^1$ Grupo de Investigaciones en Astronom\'{\i}a Te\'orica y Experimental, 
IATE, Observatorio Astron\'omico, Laprida 854, C\'ordoba, Argentina. \\
$^2$ Consejo de Investigaciones Cient\'{\i}ficas y T\'ecnicas de la Rep\'ublica 
Argentina.\\
$^3$ Department of Physics, University of Durham, South Road, Durham, DH1 3LE, UK}
\date{\today}

\maketitle 

\begin{abstract}
We use the 2dF Galaxy Group Catalogue constructed by Merch\'an \& Zandivarez
to study the large scale structure of the Universe traced by galaxy groups. 
We concentrate on the computation of the power spectrum and the two point 
correlation function. 
The resulting group power spectrum shows a similar shape to the galaxy power 
spectrum obtained from the 2dF Galaxy Redshift Survey by Percival et al., 
but with a higher amplitude quantified by a relative bias in redshift space 
of $b_s(k) \sim 1.5$ on the range of scales analysed in this work,
$0.025<k/h$Mpc$^{-1}<0.45$. 
The group two point correlation function for the
total sample is well described by a power law with correlation length 
$s_0=8.9 \ \pm \ 0.3 \ \mpc$ and slope $\gamma=-1.6 \ \pm \ 0.1$ on scales 
$s \ < \ 20 \ \mpc$. 
In order to study the dependence of the clustering properties 
on group mass we split the catalogue in four subsamples defined by different 
ranges of group virial masses. Our results are consistent with a $40\%$ 
increase of the correlation length $s_0$ when the minimum mass
of the sample increases from
$ {\cal M}_{vir} > 5 \ \times \ 10^{12} \ h^{-1} M_{\odot}$ to
$ {\cal M}_{vir} > 1 \ \times \ 10^{14} \ h^{-1} M_{\odot}$.
These computations allow a fair estimate of the relation described by the 
correlation length $s_0$ and the mean intergroup separation $d_c$ 
for galaxy systems of low mass. Our results show that an empirical scaling 
law $s_0=4.7 \ {d_c}^{0.32}$ provides a very good fit to the results from 
this work, as well as to previous results obtained for groups and clusters 
of galaxies.  The same law describes the predictions for dark matter haloes 
in N-body simulations of $\Lambda$CDM models.
We also extend our study to the redshift space distortions of galaxy groups,
where we find that the anisotropies in the clustering pattern of the
2dF group catalogue are consistent with gravitational instability, with a 
flattening of the redshift-space correlation function contours in the 
direction of the line of sight. The group pairwise velocities found from 
this analysis for a sample of groups with masses 
${\cal M}_{vir} > 5 \ \times \ 10^{12} \ h^{-1} M_{\odot}$,
are consistent with 
$\langle w^2 \rangle^{1/2} = (280^{+50}_{-110})$km/s,
in agreement with $\Lambda$CDM cosmological simulations.  The
bias factor for the 2df groups of moderate masses
${\cal M}_{vir} > 2 \ \times \ 10^{13} \ h^{-1} M_{\odot}$
is consistent with the values predicted by the combination of a CDM model
and the ellipsoidal collapse model for the formation of structures.
\end{abstract}


\section{Introduction} 
The position of groups of galaxies in the structure hierarchy makes them 
very interesting objects, as the density fluctuations sampled by groups 
lay between those traced by galaxies and clusters of galaxies. 
This is what makes the study of galaxy groups a key area of research in 
cosmology and galaxy formation. Much effort has been devoted to the study of 
groups in order to understand the large scale structure of the Universe 
(e.g., Jing \& Zhang 1988, Maia \& da Costa 1990, Ramella, Geller \& Huchra 
1990, Trasarti-Battistoni, Invernizzi \& Bonometto 1997). 
The main tool adopted in most of these works for the study of the spatial
distribution of groups is the redshift space two point correlation function 
$\xi(s)$. These studies show, as expected from the hierarchical model, that 
the amplitude of the group $\xi(s)$ falls roughly between the corresponding 
amplitudes of correlations obtained from galaxies and cluster of galaxies.
Using a suitable statistical sample of groups, Merch\'an, Maia \& Lambas
(2000) analysed the dependence of the two point correlation function 
on mass, by splitting the group sample in different ranges of virial mass, and
extending this analysis to lower mass systems (See also Bahcall \& West, 1992,
Croft et al. 1997, Abadi, Lambas \& Muriel, 1998, Borgani et al. 1999,
Collins et al. 2000).
Their results show that the amplitude of $\xi(s)$ tends to 
increase significantly with the mass of the sample. They measure a $50\%$ 
increase in correlation length for a sample with mean inter-group separation 
$d_c\simeq 28 h^{-1}$Mpc with respect to a sample with $d_c=9h^{-1}$Mpc.

On the other hand, the measurement of the power spectrum of density
fluctuations, a widely used statistical tool for galaxies and clusters of 
galaxies, has seldom been applied to the study of galaxy groups.  This
could be due to the fact that up to now, a statistically reliable
sample of galaxy groups, from which a robust determination
of the power spectrum can be made, has not been available.

Other important constraints on cosmological models can also be obtained from 
the study of the anisotropies in the redshift-space correlation function. 
The three-dimensional clustering information available in redshift surveys
is subject to the influence of peculiar velocities on the distance 
measurements. This induces an anisotropy into the clustering pattern, which 
can be used to determine cosmological parameters due to the fact that the
amplitude of peculiar velocities is related to the matter density, $\Omega$, 
and the linear bias parameter, $b$. This study is carried out by
measuring the apparent distortion of the clustering pattern in the two
point correlation function in redshift space.
On large scales, where the linear perturbation theory approximation for the 
growth of density fluctuation is valid, flows of mass from low to 
high density regions generate an artificial enhancement of the density 
contrast so as to produce a compression of the $\xi$ contours along the 
line-of-sight direction (Kaiser 1987). This phenomenon 
allows estimates of the parameter $\beta \simeq \Omega^{0.6}/b$ to be made.
On small scales, the main effect is to reduce the amplitude of density 
fluctuations by means of random non-linear motions in virialized regions, 
producing the ``Fingers of God" effect. In this regime, it is possible to 
obtain estimates of the one-dimensional pairwise rms velocity dispersion 
(Davis \& Peebles 1983). Such an analysis was recently applied to groups of 
galaxies identified in the Updated Zwicky Catalogue by Padilla et al. (2001). 
They found a one-dimensional pairwise rms velocity dispersion for groups of 
$250 \pm 110 $km \ s$^{-1}$ and a noisy estimate of the $\beta$-parameter, 
$\beta < 1.5$. These results suggest the need for larger samples of groups
in order to obtain more reliable results from these type of objects.

The 2dF Galaxy Group Catalogue (hereafter 2dFGGC), constructed by Merch\'an 
\& Zandivarez (2002) is one of the largest group catalogues at present.
These groups were identified from the 2dF public 100K data release using a
modified Huchra \& Geller (1982) group finding algorithm that takes into
account the magnitude limit of the 2dF sample, redshift completeness, and 
angular masks (Colless et al. 2001). This sample of groups has been used to
study the global effects of group environment on star formation (Mart\'{\i}nez 
et al. 2002a), the effect of local environment on galaxy member spectral types 
(Dom\'{\i}nguez et al. 2002), a statistical analysis of luminosity functions 
(Mart\'{\i}nez et al. 2002b) and a compactness analysis of groups 
(Zandivarez et al. 2002).

In this work we use the 2dFGGC as a tracer of the large scale structure 
of the Universe and compute the power spectrum of galaxy group fluctuations 
and the two point correlation function. 
The sample is being splitted by taking all galaxies above different cuts
in virial mass in order to study the dependence of the clustering strength 
on group mass. Finally, we study the redshift-space distortions of the group 
two point correlation function and obtain estimates of the group pairwise 
velocities and the group bias factors.

The layout of this paper is as follows. In section 2 we describe the group
catalogue, while in section 3 we present detailed information about 
the construction of the mock catalogues used in the clustering analysis. 
The methods for estimating the power spectrum and the two point
correlation function using the 2dF group sample are described in 
section 4. Studies of the anisotropies present in the redshift-space 
correlation function are analysed in section 5. Finally, we summarise our 
conclusions in section 6.

\section{The 2dFGGC}
Merch\'an \& Zandivarez (2002) identified galaxy groups in the 2dF public 100K 
data release of galaxies within the southern (SGP, $-37^{\circ}.5 \leq 
\delta \leq -22^{\circ}.5$, $ 21^h 40^m \leq \alpha \leq 3^h 30^m$) and 
northern (NGP, $-7^{\circ}.5 \leq \delta \leq 2^{\circ}.5$; $9^h 50^m \leq 
\alpha \leq 14^h 50^m$) strips of the catalogue (see Colless et al. 2001). 
The group finder algorithm used in the identification is an adaptation of the 
algorithm developed by Huchra \& Geller (1982), modified in order to take into 
account the incomplete sky coverage of the $100$k release of 2dF galaxies. 

The 2dFGGC was constructed using density contrast of $\delta \rho/\rho=80$ and 
a fiducial linking length velocity of $V_0=200$km s$^{-1}$ which maximise the 
group finding accuracy (see section 4 of Merch\'an \& Zandivarez 2002). 
The resulting group catalogue contains systems with at least 4 members, 
mean radial velocities in the range $900 $km s$^{-1} \leq V \leq 75000 
$km s$^{-1}$ and a total number of 2198 groups (see Mart\'{\i}nez et al. 2002b).
The limit adopted in the number of members in galaxy groups is necessary in 
order to avoid pseudo-groups.

\begin{figure}
\epsfxsize=0.5\textwidth 
\hspace*{-0.5cm} \centerline{\epsffile{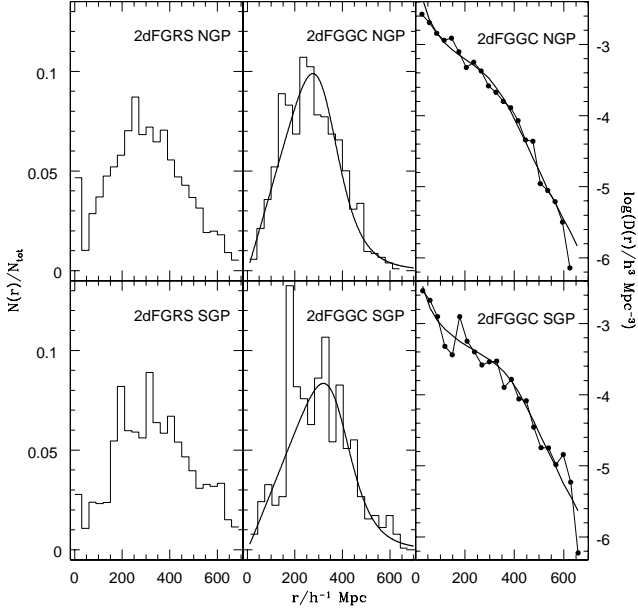}}
\caption
{Distance distributions of 2dF 100k release galaxies and groups.
The left panels show the normalised distance distributions of galaxies in the 
2dF galaxy redshift survey (2dFGRS) splitted in the northern (NGP, upper panel) 
and southern (SGP, lower panel) strip. 
The centre panels show the normalised distance distributions for the group 
catalogue (2dFGGC), while the right panels show the radial density 
distributions for the same catalogue.
The solid lines in centre and right panels show the best fitting functions
obtained for the group distance distributions (see equation 6).
} 
\label{dens}
\end{figure}

The virial group masses, ${\cal M}_{vir}$, are estimated using virial radii, 
$R_V$, and velocity dispersions, $\sigma$, ${\cal M}_{vir}=\sigma^2 R_V/G$.
The $R_V$ are computed using the projected virial radius and the $\sigma$
using the virial radius along the line of sight (see section 5 in Merch\'an
\& Zandivarez 2002). 
A robust estimation of this component is obtained by applying the bi-weight 
estimator for groups with richness $N_{tot}\ge 15$ and the gapper estimator 
for poorer groups (Beers, Flynn and Gebhardt 1990, Girardi et al. 1993, 
Girardi and Giuricin 2000).
These methods improve the velocity dispersion estimation in terms of 
efficiency and stability when dealing with small groups. 
The group catalogue has an average velocity dispersion of
$265~ \kms$, an average virial mass of $9.1\times 10^{13} \ h^{-1} \
M_{\odot}$ and an average virial radius of $1.15 ~ \mpc$.

The analysis involving computation of comoving distances are 
carried out adopting the cosmological 
model parameters $\Omega_0=0.3$ and $\Omega_{\Lambda}=0.7$.

\section{Mock Catalogues}
Throughout this work we use a set of mock catalogues constructed from several
cosmological numerical simulations of flat, low density, cold dark matter 
universes. We perform these simulations using the gravity part of the Hydra 
N-body code developed by Couchman et. al (1995), with $128^3$ particles in a 
cubic comoving volume of 180 $h^{-1}$ Mpc on a side starting at z=50.
The adopted cosmological model corresponds to a universe with a present 
day matter density $\Omega_m=0.3$, vacuum energy density $\Omega_{\Lambda}=
0.7$, baryon density $\Omega_{b}=0.0$, spectral slope $n=2$, $\Gamma=0.21$, 
Hubble constant $H_0=100 h$km s$^{-1}$ Mpc$^{-1}$ with $h=0.7$, and an 
amplitude of mass fluctuations of $\sigma_8=0.9$.   
In order to reproduce the same radial distribution as in the 2dFGRS, we 
adopt a galaxy luminosity function fitted by a Schechter function with
$M_{b_J}^{\ast}-5 \log_{10}h=-19.66$, $\alpha=-1.21$, $\Phi^{\ast}=1.68\times
 10^{-2}h^3$Mpc$^{-3}$ and a model for the average k+e corrections given by the
formula 
\begin{equation}
k(z)+e(z)=\frac{z+6z^2}{1+20z^3}
\end{equation}
(Norberg et al. 2002). Consequently, we assign absolute magnitudes to the
simulation particles in order to reproduce the previous luminosity function
and compute the apparent magnitude for the particles lying within the 2dFGRS 
mask, using the equation 
\begin{equation}
m=k+e+5\log_{10}(d_L/h^{-1}{\rm Mpc})+25+(M_{b}-5 \log_{10} h).
\end{equation}
Then we apply the magnitude limited cuts using the magnitude limited mask
of the 2dFGRS.
In order to identify groups in the mock catalogues, we adopt the same values 
$\delta \rho/\rho=80$ and $V_0=200$ km s$^{-1}$ 
which maximise the group finding accuracy on the 2dFGGC. 
It should be stated that the resulting power spectra obtained with the mock 
group catalogues following the previous recipes are very good predictions of 
the observed one. 

\begin{figure}
\epsfxsize=0.5\textwidth 
\hspace*{-0.5cm} \centerline{\epsffile{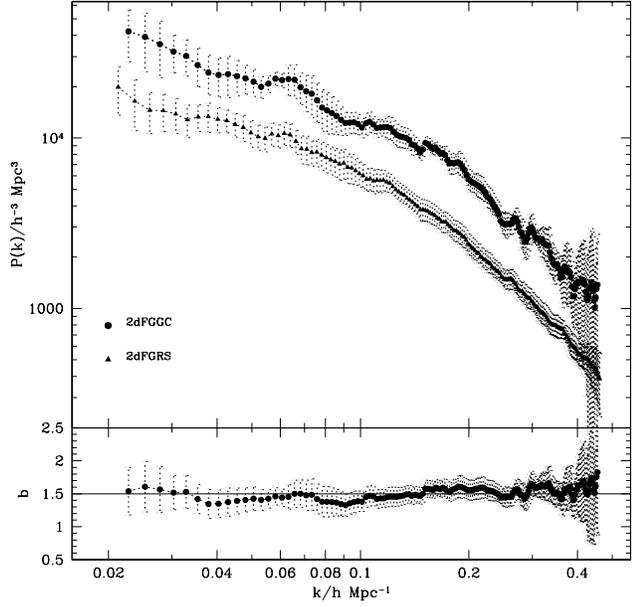}}
\caption
{Upper panel: The comparison between the power spectrum of the 
groups in the 2dFGGC (circles) and the power spectrum of the
galaxies in the 2dFGRS 100k release (triangles). The error
bars are computed measuring the dispersion over 10 mock catalogues 
constructed from N-body simulations with a $\Lambda$CDM cosmology.
Lower panel: The redshift space relative bias $b_s(k)$ between galaxies and 
groups in the 2dF survey. The error bars are computed using the usual formula 
of error propagation.
} 
\label{pkgrp}
\end{figure}

\section{Measuring the spatial distribution}

\subsection{The power spectrum}
\label{ssec:pk}
When calculating the power spectrum of the 2dFGGC we choose the 
method based on the Fast Fourier Transform (FFT) developed by 
Feldman, Kaiser \& Peacock (1994, hereafter FKP) in the version 
described by Hoyle et al. (1999).
We denote by $N_g$ the number of groups in the catalogue,
and use the vector positions ${\bf x}_c$ to specify their positions.
The following step in the FKP method is to construct an unclustered catalogue 
with the same geometry and radial selection function than the group sample.
In order to minimise the contribution of Poisson errors to our uncertainties,
this random catalogue should have a large number of points $N_r$, 
with vector positions ${\bf x}_r$. We define a quantity with mean value zero as:
\begin{equation}
\delta({\bf k})= \sum^{N_c}_{c=1} \omega(x_c) \ e^{i {\bf k} . {\bf x}_c }
- \alpha \sum^{N_r}_{r=1} \omega(x_r) \ e^{i {\bf k} . {\bf x}_r }
\end{equation}
where the first term on the right hand side is the Fourier transform of the 
spatial distribution of groups, and the second term is the Fourier transform 
of the window function of the survey. Assuming Gaussian density fluctuations, 
FKP derive a weighting function
\begin{equation}
\omega(r)= \frac{1}{1+\bar{n}(r)P_w(k)},
\end{equation}
that minimises the power spectrum variance, where $\bar{n}(r)$ is the mean
radial density of the catalogue and $P_w(k)$ is the power spectrum. 
Given that in order  to compute these weights we need an estimate of the 
power spectrum, we propose a constant value for $P_w(k)$ as an initial guess
and then test the dependence of the results on this value. 
However, the result of a series of computations of the power spectrum varying 
this initial guess indicate that the assumed value of $P_w(k)$ is not critical.
A more difficult issue is the assumed mean radial density for our catalogue.
As has been stated in many works on the 2dFGRS, the $100$k public 
release of the catalogue has a completeness mask that should be taken
into account for a reliable $\bar{n}(r)$ estimation.
In the computation of the power spectrum from the 2dF galaxies,
Percival et al. (2001) dealt with this problem by adopting a mean
radial density which depends on the position on the sky. 
In order to search for a possible variation of the radial distributions with 
the choice of 2dF strip of the catalogue, we show in Figure \ref{dens} the  
normalised distance distribution of the 2dFGGC in the northern 
(middle upper panel) and southern (middle lower panel) strips. 
The left panels show the same distribution for the galaxies in the 2dFGRS. 
The radial densities for the group catalogue are plotted in the right hand 
side panels also for both strips as in previous panels.
From these histograms one can appreciate slight differences in the radial 
distribution of groups in different strips.
Thus, in order to take into account this variation, we use different mean 
densities for the two strips in the group catalogue,
\begin{equation}
\bar{n} \longrightarrow \left \{ 
\begin{array}{l}
 \bar{n}_{1}(r,\alpha,\delta) \longrightarrow NGP \\ 
 \bar{n}_{2}(r,\alpha,\delta) \longrightarrow SGP 
\end{array} \right.
\end{equation}
where the angular dependence $(\alpha,\delta)$ is related to a correction 
that accounts for the redshift completeness mask of the 2dFGRS.
This double prescription is also adopted for the construction of the 
random catalogues in order to impose more realistic radial selection 
functions. We use smooth curves which are the best fitting functions that 
describe the histograms showed in Figure \ref{dens}. 
These functions are described by
\begin {equation}
N(r)=2^{(1+x/y)}N_{max}\left(\frac{r}{r_{max}}\right)^x\left[1+\left(\frac{r}{r_{max}}\right)^y\right]^{-(1+x/y)}\end{equation}
where the parameters are estimated using a chi-square maximum likelihood method.
The fitting functions are plotted with solid lines in central and right panel
of Figure \ref{dens}.  The best fitting parameters are $x=1.02$, $y=7.13$, 
$N_{max}=71.44$ and $r_{max}=362.34 \ \mpc$ for the NGP strip, and $x=1.03$, 
$y=7.98$, $N_{max}=66.20$ and $r_{max}=413.50 \ \mpc$ for the SGP strip. 

In Equation 3 we use $\alpha=S_c/S_r$, where
$S_c=\sum^{N_c}_{c=1} \omega^2(x_c)$ and $S_r=\sum^{N_r}_{r=1} \omega^2(x_r)$
in order to recover the definition of $P({\bf k})$ given by equation (3.4.5) 
of FKP. Then, the power spectrum estimator is obtained by:
\begin{equation}
P({\bf k})= \frac{V ( |\delta({\bf k})|^2-\alpha (1+\alpha) S_r)}
{\alpha^2 \sum^{N^3}_{i=1}(|W({\bf k}_i)|^2 - S_r^{-1})},
\end{equation}
where $V$ is the volume over which periodicity is assumed.\\
Finally, assuming isotropy we compute the power spectrum estimator
averaging over spherical shells $k<|{\bf k}|<k+dk$ in $k$-space.
We compute spectral densities at multiples of the fundamental mode in order 
to avoid oversampling of the spectrum that could result in spurious features.
The computation of these quantities is carried out by embedding the 
distributions within a periodic cubic volume $V=r_{box}^3$ which is larger than
the observational sample, and is divided into $N$ cells per side with the 
spatial distribution of points (groups or random) assigned to the grid by 
means of the nearest grid point weight assignment scheme.

\begin{figure}
\epsfxsize=0.5\textwidth 
\hspace*{-0.5cm} \centerline{\epsffile{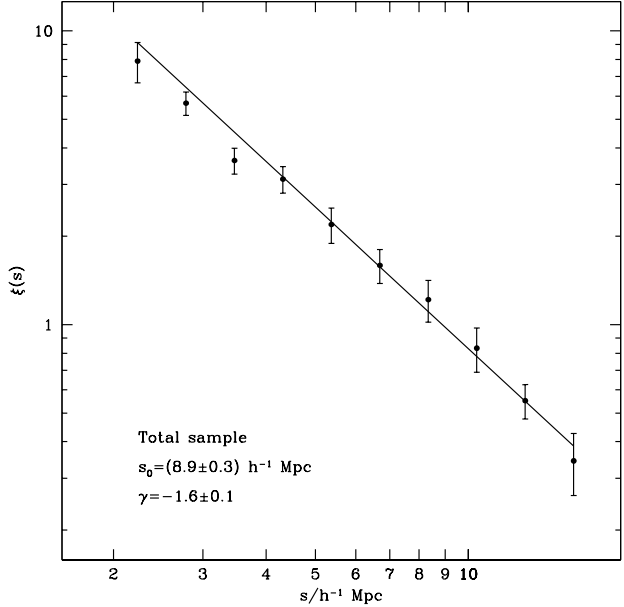}}
\caption
{The two point correlation function for groups in the 2dFGGC. The
error bars are estimated measuring the dispersion over 10 mock 
catalogues as previously applied in the power spectrum error computation.
The solid line is the best power law approximation (eq 9, see legend).
} 
\label{xi}
\end{figure}

\begin{figure}
\epsfxsize=0.5\textwidth 
\hspace*{-0.5cm} \centerline{\epsffile{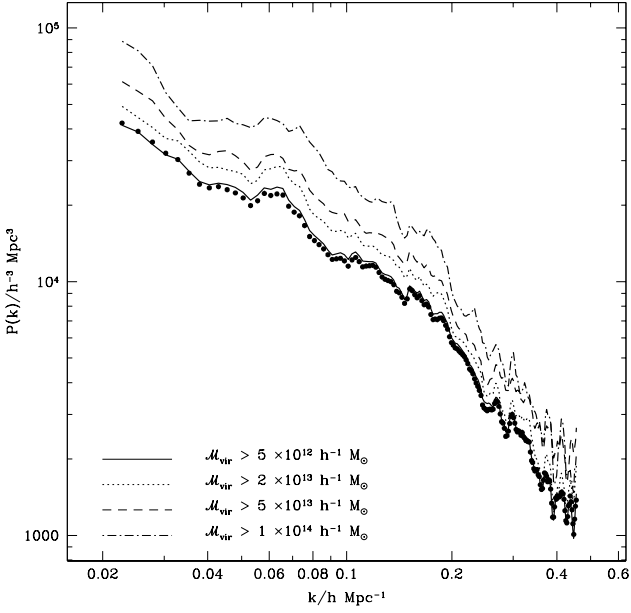}}
\caption
{The power spectrum for different mass subsamples of the 2dFGGC. 
The solid line shows the power spectrum for groups with ${\cal M}_{vir } > 
5 \times 10^{12} h^{-1} M_{\odot}$; 
dotted line shows the power spectrum for groups with 
${\cal M}_{vir } > 2 \times 10^{13} h^{-1} M_{\odot}$; the dashed line displays 
the power spectrum 
for the subsample with ${\cal M}_{vir } > 5 \times 10^{13} h^{-1} M_{\odot}$, 
and the 
dashed-dotted line shows the corresponding power spectrum for the subsample
in the range ${\cal M}_{vir } > 1 \times 10^{14} h^{-1} M_{\odot}$. 
The filled circles show
the power spectrum for the whole sample.
} 
\label{pkmass}
\end{figure}

The fiducial values chosen for the parameters involved in the calculation 
of the power spectrum are the FFT grid dimension $N=256$, the side of the box,
$r_{box}=2 r_{min}$, where $r_{min}=1244 \ \mpc$ is the side of the minimum 
box that contains the total catalogue (i.e., Nyquist frecuency $\sim 0.32
\ \up$), a constant value of the power spectrum 
for the weight function of $P_w(k)=20000 \ \up$, and a number of random points 
of $N_r=1 \times 10^6$. The resulting power spectrum for the 2dFGGC is plotted 
with filled circles in the upper panel of Figure \ref{pkgrp}. The error bars 
of this estimate are computed using a set of 10 mock catalogues constructed as 
described in previous section. We also estimate the power spectrum for the 
galaxies in the 2dF survey with fiducial parameters $N=512$ and 
$P_w(k)=5000 \ \up$, obtaining very similar results to those obtained by 
Percival et al. (2001). Our estimate of the galaxy power spectrum is shown by 
filled triangles in the upper panel of Figure \ref{pkgrp}, where the error 
bars are also estimated from mock catalogues.
As expected (Padilla \& Baugh 2002), the power spectrum of galaxy groups has 
a very similar shape to the result for galaxies but with a higher 
amplitude. This difference in amplitude can be measured computing the relative
bias between the two power spectra defined by
\begin{equation}
b_s(k)=\sqrt{\frac{P_{grp}(k)}{P_{gal}(k)}}.
\end{equation}
The bias function is plotted in the lower panel of Figure \ref{pkgrp}
where the error bars are computed from error propagation. It can be seen that 
the redshift-space bias parameter is almost constant $b_s(k) \sim 1.5$
for the full range of wave numbers, $0.025<k/h$Mpc$^{-1}<0.45$.
It should be taken into account that this redshift-space bias function could
differ from the real-space bias function. A possible relation between them
will be discussed in section 5. 

\begin{figure}
\epsfxsize=0.5\textwidth 
\hspace*{-0.5cm} \centerline{\epsffile{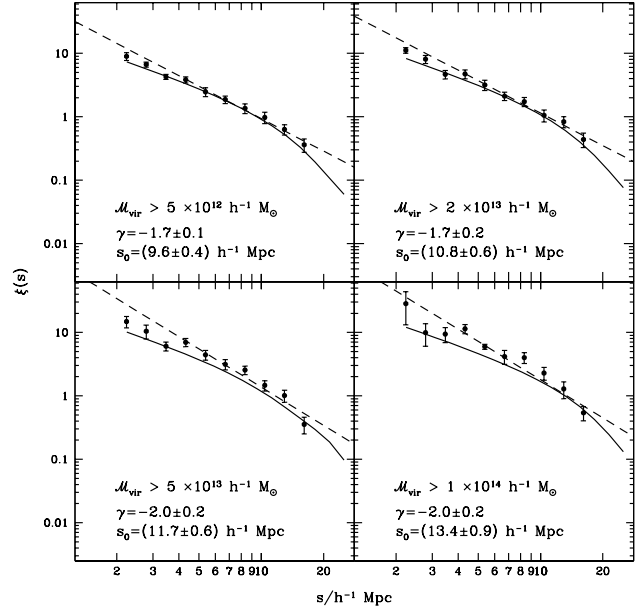}}
\caption
{The two point correlation function for different mass subsamples of the
2dFGGC. The filled circles show the two point correlation function
computed using the Landy \& Szalay (1993) estimator (see equation 9).
The dashed line corresponds to the best power law fit to the points 
(see labels), and the solid lines show the two point correlation function 
obtained from the integration of the corresponding power spectrum 
(see equation 11).
} 
\label{ximass}
\end{figure}

\begin{figure}
\epsfxsize=0.5\textwidth 
\hspace*{-0.5cm} \centerline{\epsffile{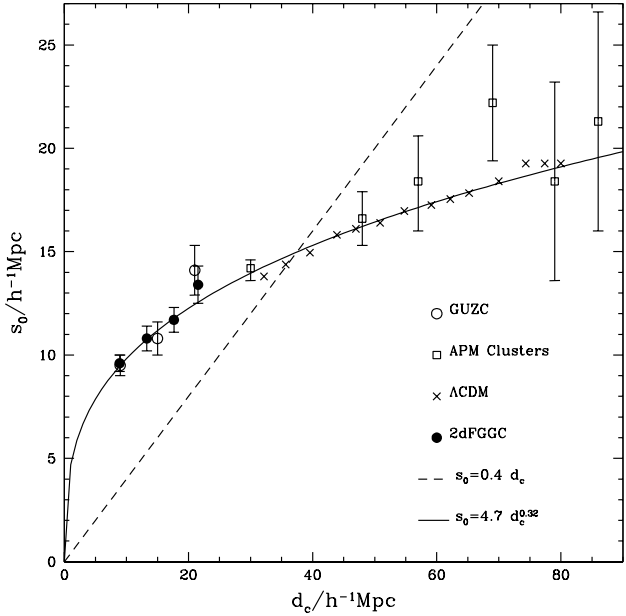}}
\caption
{Correlation length $s_0$ as a function of the mean intergroup separation
$d_c$. Filled circles correspond to the four subsamples defined in the 2dFGGC 
for different ranges of virial mass. Open circles show the 
determinations for subsamples of groups in the GUZC
(Merch\'an, Maia \& Lambas 2000). 
The open squares show the $s_0-d_c$ relation of APM clusters, and
crosses correspond to the prediction for a $\Lambda$ CDM model,
both determined by Croft et al. (1997). The dashed line represents the 
universal 
scaling law, $s_0=0.4 \ d_c$ (Bahcall \& West 1992). The solid line is our
fit to the data, $s_0=4.7 \ {d_c}^{0.32}$.
} 
\label{r0dc}
\end{figure}

\subsection{The two point correlation function}
\label{ssec:xi}

As a second method to measure the clustering properties of the galaxy 
groups in the 2dFGGC we estimate the redshift-space correlation function.
This computation is carried out using the estimator proposed by
Landy \& Szalay (1993)
\begin{equation}
\xi(s)=\frac{DD-2DR+RR}{RR},
\label{eq:xiestimator}
\end{equation}
where $DD$, $DR$ and $RR$ are the suitably normalised number of data-data, 
data-random and random-random pairs respectively, in each separation bin. 
To estimate the two point correlation function, we generate a catalogue
of randomly placed points with the same angular 
and radial selection function as the real data, as
used for the computation of the 
power spectrum.  We also take into account the completeness of the survey 
as a function of the position on the sky.

The resulting two point correlation function for groups in the 2dFGGC, 
plotted in Figure \ref{xi}, shows a positive signal up to $20 \ \mpc$.
We adopt a power law approximation for our estimate described by the
formula
\begin{equation}
\xi(s)=\left( \frac{s}{s_0}\right)^{\gamma}
\end{equation}
where $s_0$ is the correlation length and $\gamma$ the logarithmic
slope of the correlation function.
From a Levenbergh-Marquardt method (Press et al. 1986) which takes into 
account errors and applies a minimum nonlinear least-squares, the best-fitting 
parameters obtained for our estimation are $s_0=8.9 \pm 0.3 \ \mpc$ and 
$\gamma=-1.6 \pm 0.1$.
This result is consistent with previous estimates obtained
from samples with comparable space densities, by 
Girardi, Boschin \& da Costa (2000) ($s_0=8 \pm 1 \ \mpc$;
$\gamma=-1.9 \pm 0.7$) and Merch\'an, Maia \& Lambas (2000) 
($s_0=9.0 \pm 0.4 \ \mpc$; $\gamma=-1.67 \pm 0.09$).   

To test the dependence of the clustering of the 2dFGGC on the 
space density of our group samples,
we study a set of subsamples defined by different 
ranges of virial masses which translate into values of mean
inter-group separations, $d_c$.
The adopted mass limits are
\begin{eqnarray*}
Sample \ 1 & \longrightarrow & {\cal M}_{vir} > 5 \ \times \ 10^{12} \ h^{-1} M_{\odot}\\
Sample \ 2 & \longrightarrow & {\cal M}_{vir} > 2 \ \times \ 10^{13} \ h^{-1} M_{\odot}\\
Sample \ 3 & \longrightarrow & {\cal M}_{vir} > 5 \ \times \ 10^{13} \ h^{-1} M_{\odot}\\
Sample \ 4 & \longrightarrow & {\cal M}_{vir} > 1 \ \times \ 10^{14} \ h^{-1} M_{\odot}
\end{eqnarray*}
The analysis of these subsamples is carried out as follows.
We compute the power spectrum for each subsample. These 
determinations can be considered to be statistically reliable due to the 
large number of groups in the 2dFGGC. Then, we make an estimate  
of $\xi(s)$ for each subsample using the relation between the power 
spectrum and the correlation function as Fourier transform pairs:
\begin{equation}
\xi(s)=\frac{1}{2\pi^2}\int_{0}^{\infty}P(k) \ k^2 \ \frac{\sin\; ks}{ks} \ 
{\rm d}k.
\end{equation}
And finally, we compare these determinations with those obtained from the
computation of the two point correlation function for each subsample
using Equation 9. 
In Figure \ref{pkmass} we display the power spectra obtained for each
subsample. The filled circles in this figure correspond to the power spectrum
for the full sample of groups. As expected, the amplitude of the power spectrum
increases with the average mass of groups in the subsamples.
If we are to estimate $\xi(s)$ using the determinations of $P(k)$ presented
in this section, it should
be taken into account that a reliable computation requires knowledge
of the power spectrum
over a wide range of $k$ values. Since we only obtain $P(k)$ for 
$k \lsim 0.4 \ \uk$, we extend our estimates 
to larger wavenumbers assuming a power law behaviour $P(k) \propto k^{-2}$.
Consequently, we fit this power law to each subsample using
values of $P(k)$ in the range
$0.16 \lsim k \lsim 0.4 \ \uk$. The resulting power laws
obtained from a minimum nonlinear least-squares analysis are
\begin{eqnarray*} 
Sample \ 1 & \longrightarrow & P(k)= 2.37 \ k^{-2} \ \ h^{-1} {\rm Mpc}\\
Sample \ 2 & \longrightarrow & P(k)= 2.42 \ k^{-2} \ \ h^{-1} {\rm Mpc}\\
Sample \ 3 & \longrightarrow & P(k)= 2.51 \ k^{-2} \ \ h^{-1} {\rm Mpc}\\
Sample \ 4 & \longrightarrow & P(k)= 2.57 \ k^{-2} \ \ h^{-1} {\rm Mpc}
\end{eqnarray*} 
With these extensions to large wavenumbers the two point correlation
functions obtained applying Equation 11 to the results from
each sample are plotted with solid
lines in Figure \ref{ximass}. In this figure, the filled circles correspond
to the direct estimate of $\xi(s)$ for each subsample using Equation 9.
The error bars are computed using the set of 10 2dFGGC mock catalogues
described in section 3.
From this plot we observe that there is a good agreement 
between both methods of estimating $\xi(s)$.
The dashed lines in each panel correspond to the best fit obtained for 
$\xi(s)$ assuming a power law shape (eq. 10).
The best fitting parameters obtained for each subsample
are shown in table 1.

Using the estimates of $s_0$ for each subsample,
we proceed to study the $s_0-d_c$ relation. The mean intergroup separations,
$d_c$, are computed using the analytical mass function prediction based on
the ellipsoidal collapse model 
of overdensities derived by Sheth \& Tormen (1999) 
which is a very good fit to the 2dFGGC mass function
measured by Mart\'{\i}nez et al. (2002b). 
We compute the abundance $n(>{\cal M})$ of these systems in the ranges of 
masses defined for each subsample and estimate $d_c=n^{-1/3}$.

The resulting values of mean inter-group separation
for the 2dFGGC subsamples are shown in table 1.
The corresponding ($d_c,s_0$) pairs are shown in Figure \ref{r0dc} by
filled circles. For comparison, we also plot previous determinations
from groups in the Updated Zwicky Catalogue (GUZC) derived by 
Merch\'an, Maia \& Lambas (2000) which shown a very similar behaviour 
with our results.
The $s_0-d_c$ pairs shown in the range of scales corresponding to
clusters of galaxies are those obtained from
the APM cluster survey by Croft et al. (1997,
open squares). In previous works, some authors (Bahcall \& West 1992, 
Bahcall \& Cen 1992) have argued 
that Abell clusters are consistent with a linear 
$s_0-d_c$ relation described by $s_0=0.4 \ d_c$. This scaling law is
plotted as a dashed line in Figure \ref{r0dc} and does not seem to 
provide a good description either of cluster data or of groups of galaxies. 
This result was also reached by Croft et al. (1997) 
where they also computed the $s_0-d_c$ relation for clusters of galaxies
in a $\Lambda$CDM N-body simulation. 
Their results are 
represented by crosses in Figure \ref{r0dc}.
As can be seen, our estimate of the $s_0-d_c$ relation for groups 
is an extension of the relation obtained by Croft et al. (1997) from 
numerical simulations from a $\Lambda$CDM model.
Similar trends in $s_0-d_c$ were also found in other CDM models,
but with slightly different amplitudes of correlation length
(see Governato et al. 1999).
We find that a simple fit of the form
\begin{equation}
s_0=4.7 \ {d_c}^{0.32}
\end{equation}
provides a satisfactory empirical description of
the observational data as well as 
the N-body simulation results.
This empirical scaling law is shown by the solid line in Figure \ref{r0dc}.

\section{Redshift space distortions}

In this section we study the redshift-space correlation function
calculated in the directions parallel and perpendicular to the line of
sight, $\xi(\sigma,\pi)$.
This approach has led to the quantification of characteristics of the
redshift-space distribution of galaxies and clusters of galaxies, such as 
the ``fingers of God", which are elongated structures seen in redshift
surveys, originating from the random motions of galaxies inside clusters.  
The statistics that can be derived from the
anisotropies in $\xi(\sigma,\pi)$ are presented in section
\ref{ssec:zstats} and the results found for the different 2dFGGC
subsamples can be found in section \ref{ssec:zresults}.  Section
\ref{ssec:zw12} contains the results for pairwise
velocities corresponding to the different subsamples, and
\ref{ssec:zbias} the corresponding bias factors.

\subsection{Statistics from the anisotropies in the redshift-space 
correlation function}
\label{ssec:zstats}
 
The quantification of the ``fingers of God" effect comes from the 
measurement of the pairwise velocities of galaxies, which can be done
by comparing a theoretical expression for the correlation 
function, $\xi^p(\sigma,\pi)$, with the measured
values, $\xi^o(\sigma,\pi)$ (the index $o$ indicates quantities measured
from observational data, and the index $p$ denotes predicted quantities).  
Specifically, we compare curves of equal correlation function amplitude 
($\xi^{fix}$) parametrised by $r^o_{\xi^{fix}}(\theta)$ and 
$r^p_{\xi^{fix}}(\theta)$, 
where $\theta$ is the angle measured from the direction perpendicular 
to the line of sight (Padilla et al. 2001), such that 
$\xi^{x}(\sigma_r,\pi_r)-\xi^{fix}=0$,
where $\sigma_r=r^{x}_{\xi^{fix}}(\theta)\cos(\theta)$, 
$\pi_r=r^{x}_{\xi^{fix}}(\theta)\sin(\theta)$, and
the index $x$ indicates either observed ($o$) or predicted
quantities ($p$).

The predicted correlation function is obtained by
convolving the real-space correlation function, $\xi(r)$ 
with the pairwise velocity distribution function, $f(w)$, following
Bean et al. (1983).  In order to do this, we calculate
\begin{equation}
1+\xi^p(\sigma,\pi)=
\int_{-\infty}^{\infty} [1+\xi(r)] f[w' - w^s(r,r')]{\rm d}w',
\label{eq:pred}
\end{equation}
where $r^2=r'^2+\sigma^2$, and $H_0$ is the Hubble constant, $r'=\pi - w'/H_0$ 
(the prime denotes the line-of-sight component of a vector quantity) and 
\begin{equation}
w^s(r,r') \simeq -H_0 \beta \xi(r) (1+\xi(r))^{-1}r',
\label{eq:streaming}
\end{equation}
is the mean streaming velocity of galaxies at separation $r$.  

Following the usual prescription, we calculate 
the best-fit rms peculiar velocity,
$\langle w^2 \rangle^{1/2}$, for an exponential distribution, 
\begin{equation}
f(w)=\frac{1}{\sqrt{2}\langle w^2 \rangle^{1/2}}\exp 
\left( -\sqrt{2} \frac{|w|}{\langle w^2 \rangle^{1/2}} \right).
\label{eq:fw}
\end{equation}
We adopted this pairwise velocity distribution as it has shown 
to be the most accurate fit to the results from numerical simulations
(Ratcliffe et al. 1998, Padilla \& Baugh 2002).

In this work we use a non-linear CDM correlation function, for
which we assume a given cosmology.  In order to do this, we follow
the recent results from satellite and 
balloon borne cosmic microwave background experiments
such as WMAP (Spergel et al. 2003) and Boomerang (Netterfield et al. 2002), 
and assume a $\Lambda$CDM
cosmology with mass density $\Omega_{0}=0.28$, baryon density $\Omega_b=0.05$,
vacuum energy density $\Omega_{\Lambda}=0.67$,
and a CDM shape parameter $\Gamma=0.18$, and $\sigma_8=0.9$.  
These parameters are compatible with constraints from CMB data.  Even
though we 
include explicitly a baryon fraction in our models, the high level 
of noise of our observational data allow us only to measure the damping
in power produced by baryons.  This means that our choice of
$\Gamma$ implicitly allows for a combination of baryon fraction and
a real value of $\Gamma$ as discussed in Eisenstein \& Hu (1996).  
We obtain the real-space 
correlation function by Fourier transforming the CDM power spectrum
\begin{equation}
\xi^{CDM}(r)= \frac{1}{2\pi^2} b^2 \int_0^{\infty} P(k) 
\frac{\sin(kr)}{kr} k^2 {\rm d}k,
\label{eq:xicdm}
\end{equation}
and then use this to evaluate the theoretical prediction for $\xi^p(\sigma,\pi)$
using equation (\ref{eq:pred}).  
Here we assume a constant, scale independent, bias 
between the distribution of groups and mass, which has been
shown to be a good approximation for groups and clusters of galaxies
(Padilla \& Baugh, 2001).  Furthermore, by looking at figure
\ref{xi}, it can be seen 
that the values of correlation function slopes, $\gamma$, for
the different samples are not significantly different (less than
$1.3 \sigma$ away); also, figure \ref{pkgrp} shows that a constant relative
bias is a good approximation up to scales $k<0.45 h/$Mpc, equivalent
to $x \sim 10 h{^-1}$Mpc, where our analyses take place.

We search for the optimum values of the scale independent bias parameter 
$b$ and $\langle w^2 \rangle^{1/2}$ by minimising the quantity 
$\chi^2(\langle w^2 \rangle^{1/2},b)$,
\begin{equation} 
\chi^2(\langle w^2 \rangle^{1/2},b) = 
\sum_{i} [r^o_{\xi^{fix}}(\theta_i)-r^p_{\xi^{fix}}(\theta_i)]^2,
\label{eq:chi}
\end{equation} 
where we have chosen to compare a set of discrete levels, 
$\xi^{fix}=0.6,0.8,1.0,1.2$ and $1.4$, of the redshift-space 
correlation function $\xi(\sigma,\pi)$ amplitude (Padilla et al. 2001) instead
of comparing values of the correlation function on a grid of $\sigma$ and 
$\pi$ distances as done previously in other works (see for instance
Ratcliffe et al. 1998).

\subsection{Measurements of 2dfGGC $\xi(\sigma,\pi)$}
\label{ssec:zresults}

\begin{figure*}
{\epsfxsize=17.5truecm \epsfysize=4.7truecm 
\epsfbox[1 1 650 181]{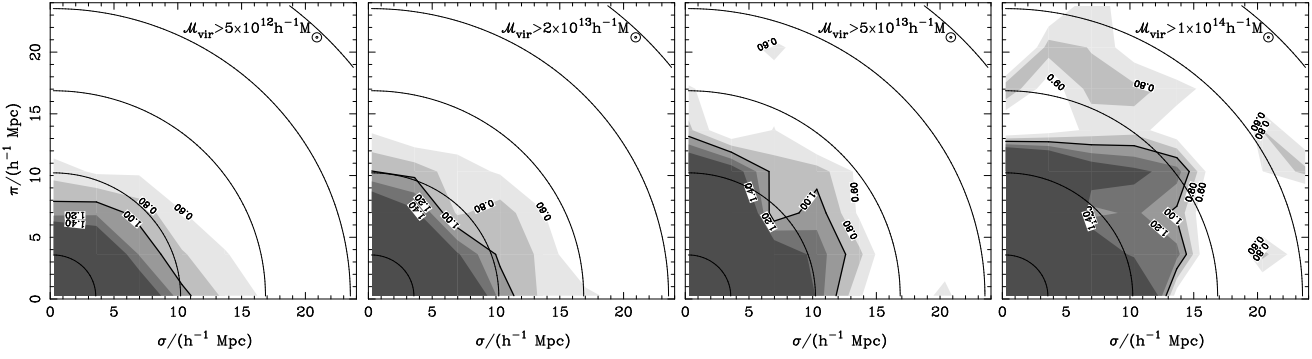}}
\caption{
The 2-point 
correlation function of the four subsamples of 2dF groups studied in
this paper,
in the coordinates $\sigma$ and $\pi$ (subsamples $1$ to $4$ from
left to right hand side panels).
The transitions between different 
shadings correspond to fixed values of $\xi=0.6,0.8,1.0,1.2,$ and $1.4$ 
levels, which are the values used later to infer the pairwise
velocities, $\langle w^2 \rangle^{1/2}$. 
The thick line corresponds to the $\xi=1$ level, and the thin lines
show the expected contours in the absence of peculiar velocities. 
}
\label{fig:dist_4}
\end{figure*}

Using the estimator represented by 
equation (\ref{eq:xiestimator}), we measure the redshift-space
correlation function of the 2dFGGC for the four subsamples
studied in this paper.  In this case, the 
quantities $DD$, $RR$, and $DR$ depend upon the separations parallel and
perpendicular to the line of sight.
The results can be seen in Figure \ref{fig:dist_4} where
the different panels show the correlation function for subsamples 1,2,3 and
4 (left to right hand side panels respectively).

Concentrating on the anisotropies we observe in this figure, 
we find that the infall pattern expected from gravitational instability
is seen in most of the group subsamples with the exception of subsample $3$. 
Here we should bear in mind the
increasingly noisy results that are obtained for the samples of higher
virial masses.  This is a simple consequence of the ever smaller number of 
more massive objects that adds a Poisson error of increasing magnitude to the
estimate of the correlation function. 
Still, subsamples $1$, $2$
and $4$ show the infall pattern for many contour levels.  

By inspecting the correlation lengths depicted by the thick
solid lines in the panels of Figure \ref{fig:dist_4},
the increase in correlation length for
subsamples with higher low virial mass limits is noticeable.  
This is in agreement
with the direct measurements of redshift-space
correlation function performed
in section \ref{ssec:xi}. 

\begin{table*}
\begin{center}
\caption{Statistical results for the 2dFGGC.  The first columns
show several characteristics of the different 2dFGGC samples, including the
minimum virial mass of the groups in each sample, 
number of groups, their average distance,
the average number of galaxy members identified per group, and the
average internal velocity dispersion as measured from the identified
member galaxies.  Columns $7$ and $8$
show the correlation function fit parameters, column $9$ 
contains the values of mean inter-group separation, and
the last two columns, the corrected values of 
group pairwise velocity and bias
factors obtained from the redshift-space distortions analysis.}
\begin{tabular}{lcccccccccc}
\hline
\hline
Sample & ${\cal M}_{vir}$ min.&$N_{grp}$ & $\langle zc \rangle$ & $ n_{member}$ & $\sigma_v$&       $s_0$  & $\gamma$ &  $d_c$   & $\langle w^2 \rangle_c^{1/2}$ & $b^{\xi}_r$ \\
       &$[h^{-1}M_{\odot}]$        &         &[$\kms$]&             & [$\kms$]  &     [$\mpc$] &          & [$\mpc$] &    [$\kms$]   &  \\
\hline
All      &                     & $2198$ &$---$    & $---$       & $265\pm150$&  $8.9\pm0.3$ & $-1.6\pm0.1$ &  $---$  & $---$&  $1.82 \pm 0.36$    \\
Sample 1 &$5 \ \times \ 10^{12}$&$1996$ &$30333$  & $6.9\pm 4.6$& $283\pm149$&  $9.6\pm0.4$ & $-1.7\pm0.1$ &  $8.94$ & $280^{+50}_{-110}$ & $1.92\pm 0.38$\\
Sample 2 &$2 \ \times \ 10^{13}$&$1448$ &$33035$  & $7.5\pm 5.0$& $333\pm143$&  $10.8\pm0.6$ & $-1.7\pm0.2$ & $15.00$ & $395^{+35}_{-95}$ & $2.04\pm 0.41$\\
Sample 3 &$5 \ \times \ 10^{13}$& $920$ &$36336$  & $8.2\pm 5.8$& $396\pm141$& $11.7\pm0.6$ & $-2.0\pm0.2$ & $20.93$ & $540^{+40}_{-100}$ & $2.24\pm 0.45$\\
Sample 4 &$1 \ \times \ 10^{14}$& $540$ &$38631$  & $8.6\pm 6.4$& $464\pm142$& $13.4\pm0.9$ & $-2.0\pm0.2$ & $28.93$ & $495^{+35}_{-175}$ & $2.51\pm 0.50$\\
\hline
\end{tabular}
\end{center}
\end{table*}

\subsection{Pairwise velocities and
correlation lengths in observational cluster samples}
\label{ssec:zw12}

In order to summarise the results obtained from the study of the
anisotropies in the correlation function of
2dFGGC subsamples, we present the values of pairwise velocities 
obtained from the correlation functions shown in Figure \ref{fig:dist_4}.

We find the group pairwise velocities by minimising
equation \ref{eq:chi}.  The real space correlation function used in this
equation corresponds to a CDM power spectrum with spectral index $n_s=1.0$,
and parameters
$\Omega_{0}=0.28$, $\Omega_{\Lambda}=0.67$, a baryon density $\Omega_b=0.05$,
a CDM shape parameter $\Gamma=0.18$, and $\sigma_8=0.9$.  

\begin{figure}
{\epsfxsize=8.truecm \epsfysize=8.truecm 
\epsfbox[56 180 580 700]{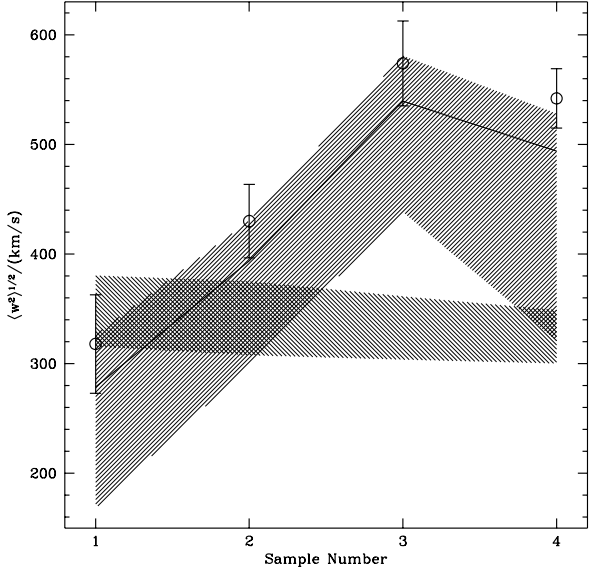}}
\caption{
Pairwise velocities obtained for
the different 2dFGGC subsamples (open symbols with errorbars).
The samples are distributed along the x-axis,
with samples of increasing lower virial mass limit towards the right.
Errorbars are obtained from the estimates of pairwise velocities
at $5$ different correlation function contour levels.
The solid line shows the corrected pairwise velocities obtained
by subtracting in quadrature the error in the distance measurement
of groups, and the corresponding 1-$\sigma$ confidence level is represented
by the upward hatched region.  The downward hatched region shows the 
1-$\sigma$ confidence level obtained from $20$ subsamples of 
groups identified from a $\Lambda$CDM simulation.  Each of these
subsamples contain the same number of groups as the 2dFGGC subsamples.
}
\label{fig:wdm}
\end{figure}

The results for the
pairwise pairwise velocities found from 
the correlation functions of the four 
2dFGGC subsamples studied here are presented
in Figure \ref{fig:wdm}.  The errorbars correspond to a $1-\sigma$
standard deviation, calculated using $5$ estimates of pairwise
velocities obtained from $5$ different correlation function level contours.
We note that this is at best a lower limit to the uncertainty
in $\langle w^2 \rangle^{1/2}$, since the results
from different $\xi(\sigma,\pi)$ can be highly correlated.
The ordering of samples is such that an increase in the $x$
direction corresponds to an increase in the sample's lower virial mass limit.

The assumption that structures form by hierarchical clustering
implies that the values of pairwise velocities found for these subsamples
should have been smaller for higher mass limits, since
more massive objects take longer to form virialized structures.  
This trend is expected in the 
direct measurements of pairwise velocities
of groups identified from a $\Lambda$CDM simulation.  
In order to test this hypothesis we take
$20$ subsets of groups with the same mass limits and number of
members as the 2dFGGC subsamples from the simulation, 
and measure their 1-dimensional pairwise velocities.
The downward hatched region in Figure \ref{fig:wdm} corresponds to the
1-$\sigma$ confidence levels for the group pairwise velocities from
the numerical simulations.  Even though the trend of decreasing
pairwise velocities for higher masses is visible in the simulation
results, the errors make it also consistent with a constant value.
As can be seen,  the agreement between subsample 1 and
the simulations is remarkable.
However, more massive groups in subsamples $2$, $3$ and
$4$ show higher values of $\langle w^2 \rangle^{1/2}$.  This behaviour
is in the opposite direction to what we expected from hierarchical
clustering arguments.  
A possible cause for this disagreement is the likely inclusion of
spurious groups affecting the low mass sample.  However,
as shown by Padilla \& Lambas (2003a), this would produce an
enhancement of the pairwise velocities, and therefore our results
for sample 1 would be overestimated, making the disagreement more
severe.  However, Padilla \& Lambas (2003a) also show that
in the case of groups identified from redshift surveys, this effect
is not important.
Another possible factor influencing our results is
the increasing values of
the average distance to groups in more massive subsamples, which
is brought by selection function effects.
The average group distance in 
subsamples 1,2,3, and 4, is larger for the most massive samples,
as can be seen in Table 1.
It is possible that
the increase in average distance translates into
a larger distance measurement error, which in turn tends to
enhance the anisotropies in the correlation function along
the line of sight, thus explaining the results obtained here.  It
should be noted though, that the amplitude of  errors needed to obtain
agreement with the simulations is small.
For instance, for samples 3 and 4, a distance measurement
error of $\Delta \langle w^2 \rangle^{1/2}\simeq 450$km/s is enough to explain
the discrepancies between the results from observations and the numerical
simulation.  This error is equivalent to a $~1\%$ uncertainty in the
group distance measurements at the mean depth of these subsamples 
(see Table 1).

The origin of the uncertainty in the group distance measurement
can be related to the fact that the number of
galaxies ($n_{member}$) used in the distance determination of the groups
shows little change among the subsamples, whereas the internal
velocity dispersion of galaxies in groups ($\sigma_v$) shows
a clear enhancement (see average values of $n_{member}$ and
$\sigma_v$ in Table 1).  An expression for this uncertainty as a
consequence of the finite number of galaxy distance measurements
per group is,
\begin{equation}
\Delta (cz) = \frac{\sigma_v}{\sqrt{n_{member}}},
\end{equation}
which is based on the assumption that these galaxies were randomly selected
from a gaussian distribution in redshift-space, of width $\sigma_v$.
Using the average $n_{member}$ in the group subsamples,
we subtract $\sqrt(2) \Delta (cz)$  in quadrature from the estimate of
$\langle w^2 \rangle^{1/2}$, and show the corrected 
$\langle w^2 \rangle_c^{1/2}$ in a solid line in Figure
\ref{fig:wdm} as a function of subsample number.  
The upward hatched region shows the range covered by a 1-$\sigma$
uncertainty in $\langle w^2 \rangle_c^{1/2}$, as it results 
from considering the widths of the
distributions in $n_{member}$ and $\sigma_v$
for the individual group subsamples, and
the original uncertainty in $\langle w^2 \rangle^{1/2}$ 
(the widths in $n_{member}$ and $\sigma_v$ are
given in Table 1 as the uncertainty in these quantities).
As can be seen, the corrected values are only slightly smaller than
the original $\langle w^2 \rangle^{1/2}$.  However, the amplitude of 
the uncertainties
is noticeably increased, improving in most of the cases
the agreement with the $\Lambda$CDM simulation.  This is not
enough to obtain an agreement for subsample 3, though, which
had also failed in showing the expected infall pattern in figure
\ref{fig:dist_4}.  

We remark however that
the assumed value of $\sigma_8$ also influences our results on
$\langle w^2 \rangle^{1/2}$, 
but to a lesser extent.  Assuming
$\sigma_8=0.7$ lowers our estimates of pairwise velocity by $\simeq
10 \%$, but leaves the uncertainty unchanged.  However, we
bear in mind the fact that these uncertainties can be underestimated
due to correlations between the individual estimates of
$\langle w^2 \rangle^{1/2}$, as explained above.

Another important point to be noticed is that 
even for sample $3$ (our sample with no infall pattern), 
the obtained value of corrected 
group pairwise velocity  $\langle w^2 \rangle^{1/2}$ 
remains one of the lowest values found for systems
of galaxies (see Padilla \& Lambas 2003b for a comprehensive
analysis of several cluster samples), and is still in agreement
with the pairwise velocities found for galaxy catalogues 
$\langle w^2 \rangle^{1/2} \simeq
(450 \pm 100)$km/s (Ratcliffe et al. 1998, Padilla et al. 2001).

\subsection{Bias factors}
\label{ssec:zbias}

\begin{figure}
{\epsfxsize=8.truecm \epsfysize=8.truecm 
\epsfbox[56 180 580 700]{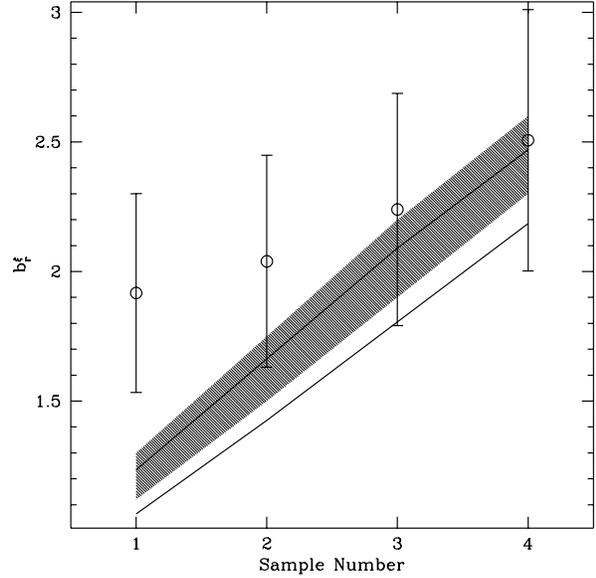}}
\caption{
Bias factors for the 2dFGGC subsamples studied in this paper.
The circles with errorbars show the results from minimising
equation \ref{eq:chi}, and the hatched region shows the range of predicted 
effective bias from CDM and the Sheth, Mo \& Tormen (2001)
mass function for $0.75 < \sigma_8 < 0.9$.  The solid lines
limit the same range, after correcting for a systematic error $\simeq 5 \%$.
The different samples are displayed along the x-axis, ordered in such a way 
as to have higher virial masses towards the right.
}
\label{fig:bias}
\end{figure}

The comparison between the measured and predicted redshift-space
correlation functions, as explained in section 2, involves the
real-space bias ($b^{\xi}_r$)
between the group and mass distributions, where the latter is
described by a CDM model.
In this work, we measure $b^{\xi}_r$ by minimising 
$\chi^2(\langle w^2 \rangle^{1/2},b^{\xi}_r) $(eq. \ref{eq:chi}), 
where $\langle w^2 \rangle^{1/2}$
corresponds to the group pairwise velocities found in the previous section.
As mentioned above, $\chi^2$ depends upon the real space correlation
function which is obtained from a CDM power spectrum
with a fixed set of cosmological parameters (see previous section).  
We minimise $\chi^2$, and show the resulting bias factors 
in Figure \ref{fig:bias} (open circles with errorbars) where the x-axis shows
the 2dFGGC subsamples in order of increasing group mass.  The errors on the bias
parameter come from the results of minimising  equation (\ref{eq:chi}) for
$5$ different levels of the correlation function.  We also include
a $20 \%$ error added in quadrature to allow for 
an uncertainty in the assumed value of $\sigma_8=0.9$ (for instance, assuming
$\sigma_8=0.7$ would produce an enhancement of a $13\%$ in our estimates
of bias factors).  The values of
bias factors and their uncertainties can be found in Table 1, where we also 
show the result obtained for the full 2dFGGC sample.  

These values can be compared to what is expected from a particular
CDM model, using the expression for an effective CDM bias as presented in
Padilla \& Baugh (2002), which is a weighted average of the
bias of a sample of clusters of mass $M$, derived
by Sheth, Mo \& Tormen (2001), and requires knowledge of the space
density of the cluster sample.  The results from using
this equation are  valid if the cluster sample
is complete above a minimum mass threshold.
The main source of systematics arising from this assumption
comes from a possible underestimation of the space density
induced by the incompleteness of the group sample.
However, a lower limit for the
completeness of the 2dFGGC subsamples is
$\simeq 0.7$, and would only induce a change of $\simeq 5 \%$
in the CDM bias factor.  

The hatched
region in figure \ref{fig:bias} shows the range of
expected values of a CDM bias factor
for the same space densities present in the
2dFGGC subsamples, assuming $0.75 < \sigma_8 < 0.9$ 
(a larger $\sigma_8$ corresponds to a lower CDM bias factor).
As can be seen, the 2dFGGC groups bias factor tends to increase with
the minimum virial mass of the sample
as expected from hierarchical clustering.  Furthermore, the
CDM bias is within the $1-\sigma$ confidence levels around 
the measured bias factors for most of the samples, the exemption being
the least massive subsample, for which the expected CDM bias factor
is smaller.  We investigated the effects of an underestimation of
the group pairwise velocities on the value of the bias, and found
that an increase of a factor $\sim 2$ in $\langle w^2 \rangle^{1/2}$
is enough for obtaining a result compatible with the CDM expectations.

As mentioned above, there is a possible $\simeq 5\%$
systematic error in the CDM bias factors.  In order to visualise the 
effects of such an error, 
we show two solid lines in figure \ref{fig:bias} which
limit the range of CDM bias factors after correcting
for this systematic error.
As can be seen, this does not significantly affect the comparison between 
the 2dFGGC and CDM model biases.  

The bias factors calculated in this section correspond
to real space quantities and are relative to the distribution
of mass which has an assumed amplitude of density fluctuations
$\sigma_8 = 0.9$.  On the other hand, the relative bias
found from the ratio between the galaxy and group power spectra
in section \ref{ssec:pk}, $b_s(k)$, 
corresponds to a redshift space quantity, and is relative to the
distribution of galaxies.  
The fact that this value corresponds
to redshift-space, allows us to make a rough estimate of what
would the corresponding real-space value be, based on
the studies of Padilla \& Baugh (2002), who find that the
ratio between real and redshift space biases for 
cluster samples is roughly $b_r/b_s =1.25$.
Assuming that this relation can be also applied to
the 2dFGGC, the redshift-space bias factor found from the
power spectrum analysis corresponds to a real-space
 $b_r(k) = 1.9 \pm 0.4$, in the range of scales studied
in this paper, $0.025<k/h$Mpc$^{-1}<0.45$, consistent with
our estimates of the real-space bias parameter from the
anysotropies in the correlation function.
Assuming a different value of $\sigma_8=0.7$ produces
no significant changes in this result.

\section{CONCLUSIONS}
We analysed the 2dFGGC constructed by Merch\'an \& Zandivarez (2002),
one of the largest group catalogues constructed at present.
Our analysis considers galaxy groups as tracers of the
large scale structure in the universe. 
The statistical tools adopted for this work are the power spectrum of density
fluctuations and the two point correlation function, both calculated
using redshift-space data. 
We calculated the group power spectrum and found that
its  shape is very similar to that of the galaxy power spectrum, but shows 
a higher amplitude as expected for higher mass systems in hierarchical
clustering.
The measurement of a relative bias between groups 
and galaxies in redshift space 
quantifies this difference in amplitude, and results 
in an almost constant relative bias value of $b_s(k) \ \sim \ 1.5$
on the scales probed by our analysis, $0.025< k/h$ Mpc$^{-1}<0.45$.

The estimate of the two point correlation function for the group sample
is found to be
well fitted by a power law of the form $\xi=(s/s_0)^{\gamma}$ with
a correlation length $s_0=8.9 \pm 0.3 \ \mpc$ and a slope $\gamma=-1.6 \pm 0.1$.

We also analyse the variation of the 2dFGGC groups clustering 
with group mass by studying the $s_0-d_c$ relation.
For this purpose, we split the group sample in four subsamples according
to different ranges of group virial masses. Firstly, we check that
the computation of the correlation function for each subsample using
the direct method defined by Landy \& Szalay (1993), yields a remarkably
similar result to that obtained by Fourier transforming the corresponding power 
spectrum estimate. Secondly, the obtained correlation length $s_0$
tends to increase when using more massive samples as previously observed
by Merch\'an, Maia \& Lambas (2000, see Table 1).

From the computation of the mean intergroup distance $d_c$, the resulting
$s_0-d_c$ relation is quite similar to that obtained by Merch\'an, Maia
\& Lambas (1993) for the GUZC. If we extend this relation with
the $s_0-d_c$ relation obtained for cluster of galaxies by Croft et al. (1997)
we observe that the universal scaling law of Bahcall \& West (1992), 
$s_0=0.4 \ d_c$, provides  a poor description to the data. 
We find that a fit described by the law
$s_0=4.7 \ {d_c}^{0.32}$, is capable of describing both, the observational
$s_0-d_c$ relation found for groups and clusters, as well as the results from
N-body numerical simulations of a $\Lambda$ CDM model 
presented by Croft et al. (1997, see also Governato et al. 1999).

The results from the study of the anisotropies in the
redshift space correlation function
show that in general, the groups in the 2dFGGC present
the expected infall pattern indicating that these systems
have not yet formed virialized structures themselves. This
can be seen from the correlation function contours of sample $1$,
which includes the majority of the 2dFGGC groups.

The study of the redshift-space correlation
functions also allowed us to make an estimate of the
groups bias factor relative to the distribution of the mass.  In order
to achieve this, it was necessary to assume a CDM model with a
fixed value of $\sigma_8=0.9$.  We compare our estimates of
bias factors to what is expected for such a cosmological model,
where the mass function is well described by 
the Sheth, Mo \& Tormen (2001) model.  We
find very good agreement for the samples $2$, $3$ and $4$.

We calculate the group pairwise velocities for the
different 2dFGGC subsamples, and correct them from
group distance uncertainty effects ($\langle w^2 \rangle_c^{1/2}$,
see table 1).
We note that the pairwise velocity measured for
samples $1$ and $2$, $\langle w^2 \rangle_c^{1/2} = (280^{+50}_{-110})$km/s and
$\langle w^2 \rangle_c^{1/2} = (395^{+35}_{-95})$km/s,
are noticeably smaller than that found for galaxies, 
$\langle w^2 \rangle^{1/2}=(450 \pm 100)$km/s,
in agreement with gravitational instability expectations.  This
agreement comes not only from a group pairwise velocity which is
qualitatively smaller than the measurements found for
galaxies, but also from a comparison with measurements
from actual simulated groups, embedded in one of the most readily motivated 
cosmological scenarios, a $\Lambda$CDM model.

\section*{Acknowledgments}
We thank Mario Abadi, Vincent Eke, and Carlton Baugh 
for helpful comments and suggestions.  We thank the Referee for
invaluable comments and advice which helped improve the previous
version of this paper.
We also thank Peder Norberg and Shaun Cole for kindly providing the 
software describing the mask of the 2dFGRS and to the 2dFGRS Team
for having made available the actual data sets of the sample.
This work has been partially supported by Consejo de Investigaciones 
Cient\'{\i}ficas y T\'ecnicas de la Rep\'ublica Argentina (CONICET), the
Secretar\'{\i}a de Ciencia y T\'ecnica de la Universidad Nacional de C\'ordoba
(SeCyT), Fundaci\'on Antorchas, Argentina, and the PPARC Rolling Grant 
at Durham.

\end{document}